%% file: main.tex
\documentclass[sigconf]{acmart}

\usepackage[utf8]{inputenc}
\usepackage{float}
\usepackage{graphicx}
\usepackage{amsfonts}
\usepackage{mathtools}
\usepackage[acronym]{glossaries}
\usepackage{enumitem}
\usepackage{adjustbox}
\usepackage{tikz}
\setcopyright{none}
\renewcommand\footnotetextcopyrightpermission[1]{}
\newif\ifcommentson
\commentsontrue

\newcommand{\new}[1]{{#1}}

\newenvironment{proofsk}{%
  \renewcommand{\proofname}{Proof}\proof}{\endproof}
\newcommand{\pbfive}{FBFT} 
\newcommand{\pbthree}{3-FBFT}
\DeclarePairedDelimiter\ceil{\lceil}{\rceil}
\begin{document}


\title{A PoW-less Bitcoin with Certified Byzantine Consensus}

\subtitle{An Unorthodox Foundation for a Central Bank Digital Currency (CBDC)?}




%



\author{Marco Benedetti, Francesco De Sclavis, Marco Favorito, Giuseppe Galano,\\
Sara Giammusso, Antonio Muci, Matteo Nardelli}
\authornote{Email address of the authors are in the form [firstname].[lastname]@bancaditalia.it, except for  giuseppe.galano2@bancaditalia.it.}
\affiliation{%
    \vspace{0.5em}
     Technical Report CFC.CRYPTO.CS/2022/1\\
    Applied Research Team (ART) - IT Department - Bank of Italy\footnotemark[2]{}
  \country{}
}


\input{abstract}

\maketitle
\pagestyle{plain}

\renewcommand{\thefootnote}{\fnsymbol{footnote}}
\footnotetext[2]{The views expressed in this paper are those of the authors and do not necessarily reflect those of the Bank of Italy.}
\renewcommand*{\thefootnote}{\arabic{footnote}}

\input{sec_intro}

\input{sec_blockchain_model}

\input{sec_pbft}

\input{sec_frost}

\input{sec_evaluation}
\input{sec_related}

\input{sec_conclusions}



\bibliographystyle{ACM-Reference-Format}
\bibliography{biblio}

\end{document}

%% file: abstract.tex
\begin{abstract}
Distributed Ledger Technologies (DLTs), when managed by a few trusted validators, require most but not all of the machinery available in public DLTs. 
In this work, we explore one possible way to profit from this state of affairs.
We devise a combination of a modified Practical Byzantine Fault Tolerant (PBFT) protocol and a revised Flexible Round-Optimized Schnorr Threshold Signatures (FROST) scheme, and then we inject the resulting proof-of-authority consensus algorithm into Bitcoin (chosen for the reliability, openness, and liveliness it brings in), replacing its PoW machinery.
The combined protocol may operate as a modern, safe foundation for digital payment systems and Central Bank Digital Currencies (CBDC).
\end{abstract}

%% file: sec_intro.tex
\vspace{-5pt}
\section{Introduction}
\label{sec:introduction}
In the past few years, the announcement of cryptoasset-inspired ``stablecoins'' by private companies (e.g., Diem by Meta\footnote{\url{https://www.diem.com/en-us/}}) and the prospective issuance by central banks of fiat currencies in digital format for retail use, or Central Bank Digital Currencies (CBDCs)~\cite{ecb2020digitaleuro, reserve2022money, chaum2021issue, cai2021blockchain}, coupled with the unabated diffusion of decentralised blockchain-based digital assets, have reignited the interest in alternative consensus protocols for blockchains, especially those amenable to permissioned settings, on which we focus here.

In this paper, we envision a distributed service provider that operates a modern, blockchain-based, programmable and transactional engine, exhibiting high-availability and strong fault tolerance.

That is a compelling motivation already, but there are far more interesting use cases, truer to the nature of a distributed ledger: Each node (or small group of nodes) may be managed by independent actors, far removed from each other, either geographically or legally.
These actors may share a common interest that would be perfectly served, technically, by a distributed ledger with no centralization point: Everyone enjoys equal rights, duties, capabilities; no one ``owns the system''; everyone contributes to its resilience.
These actors may even reside in different jurisdictions, and conform to different laws, albeit under some shared regulatory framework
\footnote{\new{One hypothetical case would be a CBDC whose high availability and fault/attack tolerance rest upon a distributed platform operated \emph{cooperatively}---in a profound sense---by several Central Banks in a given monetary area.}}.

These motivations hold for most DLTs, from Hyperledger\footnote{\url{https://www.hyperledger.org/}} to Corda\footnote{\url{https://www.corda.net/}}, and for both payment and non-payment domains. 
However, in this work, we specifically focus on Bitcoin and on \emph{digital payments}. We ask ourselves: Is the unorthodox notion of ``\emph{Precisely Bitcoin, minus its traditional consensus algorithm, plus trusted third parties, in a permissioned setting}'' a technically consistent one?

Let's start by reviewing Bitcoin and its consensus protocol.
\vspace{-5pt}
\subsection {Consensus in Bitcoin}
Bitcoin~\cite{nakamoto2008bitcoin} is a peer-to-peer monetary network launched in 2009: It implements a digital asset which does not rely on trusted third parties to guarantee its \emph{scarcity} or to prevent \emph{double spending}.
Instead of trusted parties, it employs a decentralized \emph{consensus protocol} among anonymous participants, based on Proof-of-Work (PoW). In PoW, votes (on what the next state of the system is) can be cast by just anyone, but each vote implies a substantial consumption of real-world resources (e.g., time, hardware, energy) to solve certain hard problems related to the inversion of cryptographically strong hash functions, whose solution is required to make the vote valid.
This ``costly postage stamp'' of sort is key to prevent \emph{sybil attacks} in open, anonymous settings: Without it, 
malicious actors could compromise the consensus 
by surreptitiously creating at no (or little) cost a number of pseudonymous identities, through which a majority of apparently distinct votes, hence the system, are controlled\footnote{To subjugate a PoW system, an attacker would have to outcompete the rest of the network in terms of available resources and willingness to sacrifice them. This so called 51\% attack has been widely studied in the literature~\cite{Ye2018,Saad2020,Lee2020}.}.

Many other similar crypto-assets have emerged over time, such as Ethereum~\cite{buterin2014ethereum} and Monero~\cite{alonso2020zero}, to name a few. Each of them brings in additional features (e.g., Ethereum adds a Turing-complete programming language), but for the most part they resolved to confront their large, decentralised, anonymous user base by inheriting the PoW idea made popular---and proven effective---by Bitcoin.

There is ample space for debate on whether the power-hungry PoW is inherently the best conceivable solution for large, decentralized, anonymous blockchains; perhaps the very same properties can be obtained by computationally lighter\footnote{Power-efficient alternatives for reaching a consensus in large peer-to-peer networks of anonymous participants have been explored. For example, in a Proof-of-Stake (PoS) system~\cite{King2012} such as Algorand~\cite{gilad2017algorand}, and more recently Ethereum, the voting rights are not proportional to the consumption of real-world resources but to the staking of virtual resources themselves.} means?
For sure, in a permissioned setting with few validators, PoW alone is not going to work:
The resources sufficient to outcompete a small network are likely within reach for any motivated and sponsored  attacker.

So, the question becomes:
How to disentangle Bitcoin from PoW, and by what means is the resulting blockchain supposed to keep exhibiting tolerance to faults, attacks, and censorship attempts?
%
\subsection{All of Bitcoin but PoW}
Our goal is to inherit \emph{verbatim} all the algorithms, data structures, cryptography, and software from Bitcoin, getting rid of merely the ingredients (e.g., PoW) that are unnecessary or undesirable in a \emph{permissioned setting managed by a few trusted actors}.


If it is possible to identify a small set of actors that end-users trust to cooperatively guarantee scarcity and to prevent any double spending, then a Bitcoin-like blockchain can be grown via, e.g., a consensus based on Proof-of-Authority (PoA), wherein validators are known in advance and trusted by all network stakeholders. They are ``just'' required to prove their identity by cryptographically strong means before appending any new blocks to the chain.

Of course, high availability and tolerance to faults and to malicious behaviors of some nodes (things that used to be guaranteed by decentralization and PoW) remain mandatory even in our smaller, permissioned, distributed setting. It turns out these properties can be recovered by borrowing and modifying existing consensus and signature algorithms from the literature.
The difficult thing is to inject such new algorithmic ingredients into Bitcoin while striving to maximize the reuse of its existing technical apparatus.

That's in essence the idea we develop in this paper. As usual, the devil is in the details, and it takes a lot of work to devise a ``\emph{permissioned Bitcoin}'' exhibiting all the features we call for.

The untold premise here is that there are enough virtues and strengths to be inherited from the Bitcoin codebase, even after PoW is excised, to be worth the trouble. Is this true? 
%
\subsection{Public strengths, in private}
\label{sec:bitcoin-strenghts}
Bitcoin is the one platform to combine the following characteristics.
\begin{enumerate}[leftmargin=14pt]
    \item {\bf History}. Bitcoin is to date the oldest DLT platform operating in a real-world environment. Moreover, the history of the project has shown to date a strong commitment towards the stability of the adopted technologies and long-term backward-compatibility.
    \item {\bf Focus}. The primary focal point of Bitcoin---digital payments---aligns with our prospective use cases more than other DLTs. Moreover, other non-payments use cases can also be envisioned (e.g., tokenisation, thanks to recent developments such as RGB\footnote{\url{https://rgb.tech/}}, Taro\footnote{\url{https://docs.lightning.engineering/the-lightning-network/taro/taro-protocol}}, Ordinals/BRC-20\footnote{\url{https://ordinals.com/}}).
    \item {\bf Reliability}. Bitcoin has been extensively studied by the Academy and has been open to attacks on the Internet for almost 15 years; it has had arguably more scrutiny than any other DLT. So, we reuse a wealth of battle-tested software machinery in a permissioned setting---a good starting point for, e.g, mission critical payment applications.
    \item {\bf Extensibility}. The scripting language of Bitcoin is a good trade-off between programmability and safety. It has seen the largest ever deployment of any decentralized programmable machine, while keeping a small surface of attack compared to Turing-complete DLTs. Fortunately, its programmability is strong enough to implement most (all?) second layer constructions that are relevant in our domain.
    \item {\bf Openness}. Most software/protocols in Bitcoin are open. There is a rich ecosystem of competences, software, and services around its blockchain, from open source communities and organizations---both large and small. This implies a level-playing field open to, e.g., small Fintechs, which is important in potential pro-competitive public shared platforms.
    \item {\bf Liveliness}. The communities developing Bitcoin are large, lively, and diverse; a profusion of new features/updates are always in the work. And, while the \emph{network effect} sustaining the Bitcoin stack is one of the largest in existence as far as DLTs are concerned, there is \emph{no private organisation in key roles}: A nice attribute for applications such as CBDCs.
\end{enumerate}

The most part of all these features are PoW-independent. There is much to reuse in a permissioned, payment-oriented blockchain.
%
\subsection{Solution overview}
We target settings where there are some (from $4$ to $\approx 20$) privileged and trusted nodes in charge of accepting and validating all transactions and growing an otherwise Bitcoin-like blockchain.

We analyzed several BFT consensus algorithms (e.g., \cite{GolanGueta2019,baudet2019state,Yin2019,Buchnik2020,Distler21surveyBFT}), and we decided to design our block creation process around PBFT~\cite{Castro1999}. 
PBFT sacrifices linear communication (the number of exchanged messages is quadratic in the cluster size) in return for a simpler implementation; however, our requirements call for the mining network to produce no more than a block per minute, and our cluster is (by design) small enough to make the superlinear communication complexity a minor drawback, whereas simplicity in the implementation helps a lot the cohabitation with the Bitcoin platform. We implemented the BFT consensus algorithm from scratch instead of relying on already existing blockchain implementations, such as Tendermint~\cite{Buchman2016}, in order to maximize the reuse of and compatibility with the Bitcoin protocol, including its consensus engine.

The acquisition of block signatures from a valid quorum of trusted nodes is performed by FROST~\cite{Komlo2020} (\emph{Flexible Round-Optimized Schnorr Threshold Signatures}). This is possible thanks to the recent introduction, within Bitcoin, of Schnorr signatures~\cite{Schnorr1989}, via the Taproot soft-fork~\cite{wuille2020taproot}

Unfortunately, a simple juxtaposition of PBFT and FROST is not enough: Issues arise during distributed signature, because the consequences of the possible reluctance of (faulty or malicious) nodes to sign blocks is something PBFT is unaware of and FROST alone is unable to deal with. In addition, both protocols have parameters on which the properties of the emerging system depend. These parameters have to be chosen to play nicely with each other. We work out a solution to this intermixing problem in the next sections.
%
\subsection{Scope of this work}
We focus only on the foundational issue of the consensus and signing protocol at the ``on-ledger'' layer, i.e., on designing and developing a working PoA-based BFT algorithm meant to sustain the growth of an \emph{\new{extremely}} Bitcoin-like blockchain.

True to the nature of Bitcoin, we want our solution be as open and as subject to scrutiny and reuse as possible: A prototype implementation of the entire system is available in open source\footnote{\url{https://bancaditalia.github.io/itcoin}}.

But, there is a lot more to any real-world Bitcoin-derived permissioned payment system than is dealt with in this paper. 
Two of the most pressing issues of the retail payment systems we are interested in supporting with our construction are \emph{scalability} (to the order of tens of thousands of transactions per second) and \emph{privacy} (of the payment metadata with respect to centralized parties and payment processors). Both properties are out of scope for the present work. They are meant to be achieved by \emph{programming the core distributed machine} we devise into guaranteeing them. The so called second layer protocols we will leverage to obtain this result are ongoing research and are discussed as future work (cfr. Section~\ref{sec:future}).
%
\subsection{Structure and contribution of this report}
\new{Our major contribution is to show how 3 fairly sophisticated protocols, coming from different communities---namely Bitcoin, PBFT, and FROST---can be altered to make them interlock neatly with one another. From such pooling, a permissioned Bitcoin-like DLT emerges, with strong fault tolerance and confidential aggregation of signatures.}
\new{A detailed specification and an open-source implementation are contributed.}

\new{To the best of our knowledge, this is the first time algorithms such as PBFT and FROST are combined and adapted to a PoA setting that retains the wealth of technical tools accrued by Bitcoin.}

\new{A high-level overview of the architecture and interlinks of the  system is provided in Section~\ref{sec:blockchain-model}, together with a discussion of the desired requirements for such a PoA-based mining federation.}

\new{Then, one section is devoted to each of the three protocols, with the aim of describing the adaptations and enhancements they go through in order to smoothly engage with each other:}

\begin{itemize}
\item \new{Section~\ref{sec:variations-from-bitcoin-block} reviews and characterizes the changes we apply to Bitcoin, and show how much (or how little) our flavor of the protocol differs from the public one;}
\item \new{Section~\ref{sec:pbft} describes the features and implementation of a BFT consensus algorithm modified for use by a mining network in cooperation with a suited signature algorithm; we move from a reference protocol (PBFT) and modify it to fit our architecture and requirements;}
\item \new{Section~\ref{sec:frost} presents \pbfive{} and \pbthree{}, two novel FROST-derived protocols to aggregate a quorum of block signatures into a single one (with advantages in terms of confidentiality and space efficiency) during PBFT consensus rounds.}
\end{itemize}
\new{Finally, we evaluate the proposed solution in a geographically distributed environment (Section~\ref{sec:evaluation}), we review the related literature in the areas of permissioned DLTs, custom BFT designs, and threshold signature schemes, comparing our work with the state of the art (Section~\ref{sec:related}) and we list a few extensions and future developments that would allow our new architecture to be used in real-world scenarios (Section~\ref{sec:future}).}

%% file: sec_blockchain_model.tex
\section{System Model and Requirements}
\label{sec:blockchain-model}
%
%
%
%
%
%
In this section, we describe at a high-level the system architecture and the desired properties it is expected to fulfill.
\subsection{\new{High-level architecture}}
\label{sec:architecture}
\begin{figure}[]
\centering
\includegraphics[width=0.8\columnwidth]{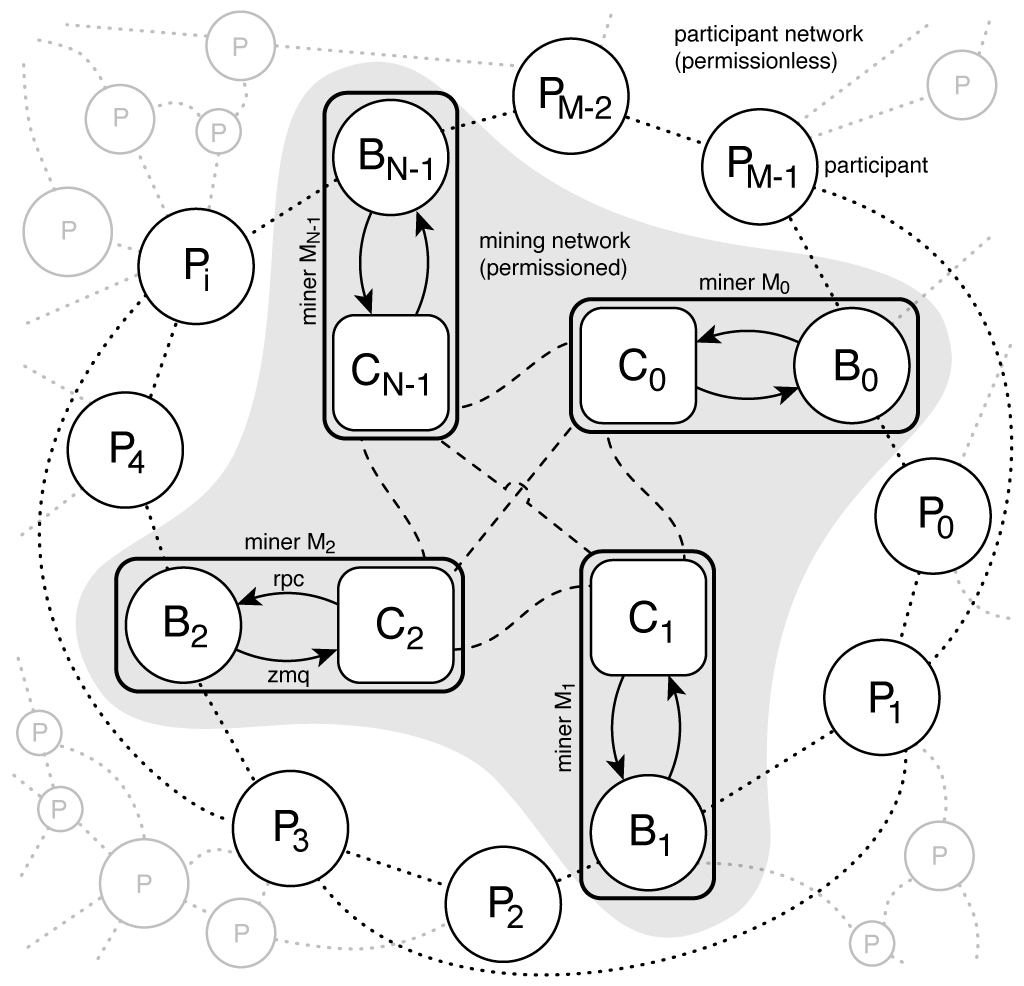}
\caption{A permission-less participant network and a permissioned mining network with $N=4$ nodes.
\label{fig:system-architecture}}
\end{figure}

Our architecture is composed of two networks with different properties: a \textit{participant network} and a \textit{mining network}---see Figure~\ref{fig:system-architecture}.

\paragraph{Participant network.} \new{The participant network is composed of a set of \textit{participant nodes}, noted $P_0, P_1, \dots, P_{M-1}$, which run the modified Bitcoin protocol from Section~\ref{sec:variations-from-bitcoin-block}.
Each participant node receives, validates, and stores a copy of our Bitcoin-like blockchain.
Participants form a spontaneous, \textit{permission-less} peer-to-peer network, without a predefined topology or size.
%
%
%
%
The bidirectional communication channels among them (dotted lines in Figure~\ref{fig:system-architecture}) are used to propagate blocks and messages via gossiping, just like in Bitcoin.}
\paragraph{Mining Node}
\new{The rounded rectangles inside the grey area are our mining\footnote{The terms ``miner'' and ``mining'' are etymologically incongruous in the context of our architecture, where trusted nodes do not operate to \emph{mining} any reward; however, we stick to them for historic reasons and for their close association with Bitcoin.} nodes, or ``miners'' (there are four of them in Figure~\ref{fig:system-architecture}). 
Each miner $M_i=(B_i, C_i)$ is composed of a \emph{bridging node} $B_i$ and a \emph{consensus node} $C_i$, running on the same host and connected by synchronous \emph{bridging channels} (see next).
%
%
%
%
Each miner is controlled and operated by one member of a federation of $N$ identifiable actors, called \emph{validators}.
While the bridging node of each miner runs the same protocol as any other participant node (in particular, it collects transactions to be validated from participants and propagates new valid blocks to others as soon as it gets aware of them), the consensus node runs the modified version of PBFT and FROST described in Section~\ref{sec:pbft} and Section~\ref{sec:frost}, respectively.}
\paragraph{Mining network.}
\new{Miners are connected to each other in a full mesh topology; the resulting \emph{permissioned} network is called the \emph{mining network} (everything within the gray area in Figure~\ref{fig:system-architecture}).
This is a peer-to-peer network too: Mining nodes are equivalent to each other, with no one playing any special role. 
The communication links among mining nodes (dashed lines) are bidirectional channels used to exchange authenticated\footnote{Each mining node has a private key that is used to sign messages; the corresponding public key are used by other mining nodes to verify the origin and authenticity of each consensus message they receive.} messages required by the PoA consensus and signing protocols (as per Section~\ref{sec:pbft} and Section~\ref{sec:frost}).}
%
%
\paragraph{Bridging channels.}
\new{In between the bridging node and the consensus node of each miner, there are 2 host-local, synchronous channels (solid, oriented arcs in Figure~\ref{fig:system-architecture}):
They act as \emph{bridging channels} between the Bitcoin realm and the PBFT/FROST one.
One uses an RPC protocol, whereby the consensus node takes the initiative to interact with the corresponding bridging component to, e.g., obtain a candidate template block, sign a block, ask to broadcast a block (see Section~\ref{sec:pbft-normal-operation} and all the self-loop messages in Figure~\ref{fig:pbft-normal-operation}).
The other bridging channel adopts a publish/subscribe model over the ZMQ protocol: The consensus node subscribes to the bridging node in order to get the mining federation notified of occurrences of new signed blocks, which act as PBFT checkpoints (Section~\ref{sec:pbft-checkpoints}).
These protocols and modes of interaction were chosen because the corresponding endpoints are already exposed by the standard Bitcoin core APIs offered by $B_i$, i.e., for maximum Bitcoin reuse.
}
\paragraph{Roles and coupling.}
\new{The mining network is a service provider: Its goal is to collect transactions from participants, reach a robust consensus on which to include in new valid blocks, and then deliver signed blocks back to participants, thus growing their shared and trusted blockchain.
Thank to the properties of the consensus and signing protocols, the mining network appears to the participants as a single mining entity.
Dually, the network of participants acts as a unique, virtual client submitting transactions to the blockchain managed by the mining nodes, and expecting such transactions to be timely validated with cryptographic strength.
The client (network of participants) is reliably connected to the server (network of miners) via a few standard Bitcoin-like $(P_j, B_k)$ channels freely established by at least some participant $P_j$ towards one or more of the bridging nodes $B_k$; these channels are indistinguishable from regular peer-to-peer channels within the permissionless network.
}
\paragraph{Failures.}
\new{We assume a Byzantine failure model where $F_B$ mining nodes can fail\footnote{A mining node fails if and only if the mining component fails, or the participant component fail, or any bridging link fails.} arbitrarily.
In addition to Byzantine failures, we assume that further $F_C$ nodes may crash.
The consensus algorithm we employ rely on synchrony to provide liveness, but not to provide safety.
To avoid the FLP impossibility result~\cite{fischer1985impossibility}, we assume that (dashed) communication channels are weakly synchronous: Message delays among correct miners do not grow too fast and indefinitely, because a Global Stabilization Time (GST) event~\cite{Dwork1988} eventually happens, after which the mining network behaves synchronously. Moreover, as in ~\cite{GKL15}, we assume that all participants are able to synchronize in the course of a ``round'', and that each round includes a GST event. This is considered a pretty faithful model of a real production system network, where faults are eventually repaired. 
As long as the network is in a failed state, it may fail to deliver messages, delay/duplicate them, or deliver them out of order.
We assume an adversary that can coordinate faulty nodes, but cannot subvert the cryptographic primitives in use.}
%
%
%
\subsection{Requirements}
%
%
%
\new{We call for our PoA consensus to exhibit the following properties.}
%
%
\begin{enumerate}[label=R\arabic*,leftmargin=*]
  \item \textbf{Correctness}. All blocks need to have a content that is valid according to the rules of the blockchain application, e.g., valid hash of the previous block, valid signatures, non-negative balances. Each block must transition the blockchain from one valid state to another, and the mining network must prevent invalid blocks to be produced.
  This property is also called \emph{validity} in the context of blockchain networks, and \emph{consistency} in the context of database systems (ACID properties).
  
  \item \textbf{Safety} (or, agreement, deterministic finality).
  In a permissioned blockchain, safety forbids chain forks, i.e., different but valid versions of the most recent blocks of the same blockchain. This requires the Common Prefix Property~\cite{GKL15} to be deterministic instead of probabilistic. Specifically, at the end of a round, if a honest participant ``prunes'' $k \ge 0$ blocks from the tip of its chain, the probability that the resulting pruned chain is not a prefix of another honest participant chain is exactly $0$ (instead of exponentially decreasing with $k$, as in PoW blockchains)\footnote{
In Bitcoin and other ``\emph{Nakamoto consensus protocols}'', transaction finality is a \emph{probabilistic} concept: 
The mere appearance of a transaction in a new block does not offer solid guarantee that it will not be (possibly maliciously) reverted. There is still the possibility that an attacker builds an alternative blockchain that "reorganizes" the latest (i.e., the longest) accepted version of the ledger to subvert or revert any transaction. However, once a transaction is included in a block, the probability that an attacker with finite resources succeeds in building such an alternate reality drops exponentially with the number of blocks appended to the chain.} and this requires the Common Prefix Property to hold $ \forall k \geq 0 $, which is equivalent to the absence of forks. 
This property is also called \emph{agreement} in the context of distributed systems, and \emph{finality} in the context of payment systems\footnote{The deterministic finality of transactions we have  regained is a feature of no little bearing on the legal status of the digital assets that may be transacted on the platform.
} 
  (e.g., Libra \cite{baudet2019state}).
  

  \item \textbf{Liveness}. Within each round, new blocks must be produced every \textit{block time} and propagated to the participants network every \textit{round time}.
  We use the Chain Growth property~\cite{GKL15},
  with parameters $\tau = \frac{\textit{round time}}{\textit{block time}} \in \mathbb{R}$ and $s \in \mathbb{N}$: For any honest participant, it holds that after any $s$ consecutive rounds it adopts a chain that is at least $\lfloor \tau \cdot s \rfloor$ blocks longer.
  
  \item \textbf{Calmness}. The pace of block production is upper-bounded, which helps participants to form expectations on their resource requirements. If a Byzantine miner creates blocks at a rate significantly higher than $\frac{1}{\textit{block time}}$, it can cause participants to run out of resources, effectively carrying out a denial-of-service attack.
  We require that after any $s$ consecutive rounds it adopts a chain that is at most $\lfloor \tau \cdot s \rfloor$ blocks longer.
 
  \item \textbf{Confidentiality}. This property requires that, at each round carried on \emph{with no faulty/Byzantine miners}, the mining network does not reveal information to the participants, other than the new block and its solution\footnote{\new{This property is impossible to guarantee in rounds where Byzantine failures happen in our architecture, since the network configuration is known to each participant, and a Byzantine node can choose to reveal extra information to the outside world.}}. In other words, the participants should not learn anything other than the fact that a provably valid new block was added. Other information, such as the miner who forged the block, the active miners/signers who approved it, the total number of miners/signers in the federation, should be kept hidden from the participants. This property makes targeted attacks against the mining network harder.
\end{enumerate}
\new{R2-R3 have been already defined and studied in the context of blockchains~\cite{GKL20}, whereas R4-R5 are somewhat peculiar to ours.}
%
\subsection{Optional requirements}
There are at least two other secondary requirements that could be taken in account, but that are out of scope for this work. One is the \emph{scalability} of the BFT solution in terms of number of messages exchanged among the mining nodes at each round, as a function of the number of mining nodes\footnote{In the literature, two properties related to scalability have been considered: \emph{linearity} and \emph{responsiveness} (e.g., see~\cite{Yin2019}).
Linearity (a condition considered optimal in the context of BFT) guarantees that creating new blocks incurs only a linear communication cost, even when leaders rotate.
Responsiveness means that the leader has no built-in delay steps and advances as soon as it collects responses from validators.}.
In this paper, and considering our target use cases and the typical size of the mining network, this notion of scalability is not of paramount importance. 
Another property is that of \emph{fairness}, i.e., 
preventing a Byzantine miner from delaying/censoring certain transactions from the network (can only happen under specific conditions). The measures to include to eliminate this specific kind of unfairness are known, but they unnecessarily complicate the baseline construction we are interested in presenting here, and are left as future work (cfr. Section~\ref{sec:future}).
%
\section{Amending the Bitcoin protocol}
\label{sec:variations-from-bitcoin-block}
This section describes the main changes to the Bitcoin protocol.
\paragraph{Block validity.}
\new{Our blocks are valid only if they include the solution to a specific ``block challenge'', as in the Bitcoin Signet~\cite{signet_bip325}. 
It could be expressed either as a Bitcoin script via \texttt{OP\_CHECKMULTISIG}, or as an aggregated public key (see Section~\ref{sec:frost}).
The challenge is distributed among participant nodes at setup time, and set in the configuration file of each node they run.
%
%
For each block, the data that satisfy the challenge, called "block solution", is stored in a special \texttt{OP\_RETURN} output of the coinbase transaction, so it is automatically propagated to the peer-to-peer network via the standard mechanisms used for blocks and transactions.}

\new{As we shall see (Section ~\ref{sec:frost}), in our case the solution represents an ``aggregated signature'', i.e., a set of signatures from a valid but opaque quorum of trusted signers in the mining network who agreed (Section ~\ref{sec:pbft}) to append that specific block at a specific height.}

\new{To accommodate for our safety requirement (R2) in the context of a PBFT-derived consensus\footnote{
\label{note:x}In presence of delays or failures, it is impossible to know in advance the specific quorum of validators that will agree to sign a block at a specific height, as it would imply foreseeing if and how failures will occur.}
(see Section~\ref{sec:pbft}), it is necessary to alter the blockchain validation rule to exclude the block solution from the computation of Coinbase transaction hash, and consequently from the Merkle root for transactions\footnote{\new{Otherwise, different sets of signers for the same block (see Note~\ref{note:x}) could lead to different, valid block solutions. This would result in different coinbase transactions and block hashes, which in turn would lead to chain forks, eventually triggering a chain reorganization.
By our safety requirement (R2), this possibility is to be ruled out.}}.}

In addition, it is necessary to include the PoW fields \texttt{nBits} and \texttt{nNonce} in the block signature. This is because we want to prevent a (malicious) miner with SHA-256 hashing power to be able to cause a fork by tweaking them: If the PoW fields were not signed, then a miner could change \texttt{nNonce} to imply more work, and its block would replace the legitimate one by the Bitcoin rules\footnote{An alternative solution to this problem would be similar to the one used with the block signatures, i.e., to exclude the PoW fields from the block hash altogether. However, this approach would call for a much more invasive modification to the existing Bitcoin code base, and this is contrary to our goal of maximizing its reuse.}.
%
\paragraph{Block mining} The steps for creating blocks become as follows:
\begin{enumerate}
  \item \new{Upon request by the consensus node of a miner, the corresponding bridging node assembles a block template}, i.e., it selects a set of transactions from the \new{mempool}, and adds a coinbase transaction with an empty block solution. At the end of this step, \emph{the block Merkle root is finalized}.
  \item The miner \textit{grinds} the block, i.e., it finds a nonce that fulfills a trivial PoW-like challenge, which is purposely included for backward compatibility with the original Bitcoin protocol. At the end of this step, \emph{the block hash is finalized}.
  \item A quorum of miners signs the block, i.e., it appends a valid block solution. At the end of this step, \emph{the transactions are finalized}: the Merkle root and block hashes are unaffected.
\end{enumerate}
\paragraph{Block interval.} 
Provided that, in our vision that we borrow from Bitcoin, retail transactions scalability is achieved off-chain, e.g., using a layer-2 Payment Channel Network, the on-chain scalability requirements can be limited, on average, to a $\approx 1.8$MB block per minute, differently from the original Bitcoin configuration, where the PoW target difficulty is adjusted periodically so to generate a block roughly every 10 minutes. Theoretically, we could lower the block time further, but this would affect the ability of the participant network to be in sync with low bandwidth requirements.

\paragraph{Block subsidy.}
Differently from the public Bitcoin, we allow any value for the block subsidy, i.e., for the freshly minted coin that is output by any coinbase transaction. This is done by removing the block subsidy checks from the code base of participant nodes\footnote{It is worth recalling that the block subsidy plays a very specific role in the public Bitcoin, i.e., to provide incentive to miners, who pay for the resources they invest in mining (and then accrue some revenue) by the market value hopefully recognized to the very subsidy tokens they mine. Especially in early times (when transaction fees play almost no role as there are few users transacting) this was an essential bootstrapping mechanism, subject to a well known ballistic halving procedure, still underway, meant to smoothly transition the system from a ``self-referential bet'' into a full-fledged and largely used service with a fee-based sustainability model. All these motivation, cost-recovering, and bootstrapping phenomena are \emph{non-existent in our permissioned setting}: They are replaced by external incentives, agreements, and monetary flows specific to the supported use case and to the specific mining federation. The block subsidy maintain one last, key role though: It is the one technical mechanism through which the asset issuer(s) in the mining federation inject freshly minted tokens/money into the blockchain, subject to, once again, \emph{external} policies and arrangements.}.

\paragraph{Coinbase maturity.} In Bitcoin, coinbase transaction outputs can only be spent after a certain number of new blocks (100 in the public network). This number is called \emph{coinbase maturity}.
Its existence is a countermeasure to rule out certain inconsistencies and disservices to end users in case of blockchain reorganizations\footnote{Without a large maturity value, the coinbase transactions of orphan blocks would become invalid in case of a reorganization, together with any subsequent transactions that depend on their outputs, causing severe inconveniences to end users.}. 
In our settings, no forks occur as per our safety requirement, and no reorganization can happen, so the coinbase maturity is safely set to 0.

\vspace{5pt}

A suggestive (if insubstantial) appraisal of how much of the public Bitcoin code we retain is obtained by measuring the syntactic scope of our changes; they amount to (a) the replacement of $\approx20$ lines of code and (b) the addition of $\approx500$ lines.
This patch is sufficient to glue the latest Bitcoin core\footnote{To fully profit from features (4) and (5) in Section~\ref{sec:bitcoin-strenghts}, our open source patch and code are always kept up to date with the latest release of the public code. At the time of writing, we are ``permissioning'' the Bitcoin core version v23.0.} to the implementation of the protocols described in the rest of this paper, and to tie up all loose ends. Such custom protocols add another $\approx8k$ lines of code, i.e., less than $2\%$ of the current Bitcoin core size.







%% file: sec_pbft.tex
\section{Ordering blocks with PBFT}
\label{sec:pbft}
Our liveness (R3) and calmness (R4) requirements call for the mining network to produce a block per minute (cfr. Section~\ref{sec:blockchain-model}). 
We assume that our technological infrastructure is up to the task in terms of computing power and network bandwidth/latency, and we design our block creation process around a Practical Byzantine Fault Tolerant (PBFT) protocol~\cite{Castro1999}.
The preference for PBFT, in lieu of other BFT algorithms, is motivated by the following consideration: PBFT sacrifices linear communication (i.e., the number of messages exchanged is not linear in the cluster size) in return for a simpler implementation; however, our cluster is small enough, by design, to make the superlinear communication complexity a minor drawback, whereas simplicity in the implementation helps a lot the cohabitation with the complex Bitcoin protocol. 
%

\subsection {PBFT in a nutshell}
PBFT is a state machine replication algorithm: it relies on a set of \textit{replicas} to maintain a service state and to implement a set of \textit{operations} onto it. 
The replicas move through a succession of configurations called \textit{views}, which are numbered consecutively. In a view, one replica is the \textit{primary} and the others are \textit{backups}. \textit{View changes} are carried out when it appears that the primary has failed. 
Service operations are invoked by \textit{clients}, which send \textit{requests} to the primary. 
Then a three-phases protocols begins, that allows replicas to agree among them on the order in which requests are to be executed: (i) in the \textit{pre-prepare} phase, the primary assigns a sequence number to the request and multicasts it request to the backups; (ii) in the \textit{prepare} phase, the backups agree on the sequence number proposed by the primary; (iii) in the \textit{commit} phase, the replica confirm that an agreement on the request and its sequence number has been reached by a \textit{PBFT quorum} of replicas. Then, each replica executes the operation and replies to the client.
The client waits for $F_B+1$ replies from different replicas with the same result, where $F_B$ is the maximum number of replicas that may be Byzantine.

\vspace{5pt}

In the following sections we describe a specialized version of PBFT that deals with block selection and block signing in presence of Byzantine and crash failures.
The algorithm contains additional blockchain-specific steps, but also some simplifications, based upon the following considerations: (i) Our state machine exposes only a single operation, that is the appending of a new block; (ii) Our mining network has a single abstract client that is the network of participants; (iii) There are no parallel mining requests, since each miner expects to mine only the next block (the one after the next block cannot be mined if the next block is not mined yet); (iv) The \emph{checkpoint} and state propagation mechanisms can rely on the block propagation process already present in the participants network.
%
\subsection{Quorum size}
\label{sec:pbft-quorum-size}
Suppose we want to tolerate, at most, $F_B$ Byzantine failures, and $F_C$ crash failures in the mining network.
We replicate the service across $N=3F_B + 2F_C + 1$ nodes, and we define \textit{Byzantine quorum} as $Q=2F_B + F_C + 1$.
For example, suppose the blockchain operators want to keep mining blocks after the private key of one node is compromised (Byzantine failure) and at the same time there is ongoing maintenance on another node (crash failure).
It is $F_B=1$ and $F_C=1$, so we need $N=6$ nodes and a PBFT quorum of $Q=4$.
This network configuration employs the minimum $N$ and $Q$ guaranteeing that (i) two different quorums always intersect in at least one non-Byzantine mining node, i.e., $ Q=\ceil{ (N+F_B+1)/2 } $, which is a necessary condition for Safety (P2) and (ii) a quorum of non-faulty miners can be reached also if $F_B + F_C$ nodes fail, i.e., $N-Q = F_B + F_C$, which is a necessary condition for Liveness (R3)---provided delays among correct nodes do not grow indefinitely. 

In addition to the PBFT quorum, we need define the \emph{Reply quorum}, i.e., the number of signatures which are required for a block to be valid before the network of participants.
Assuming that each node controls a single key, then a safe Reply quorum also depends on $F_B$. The minimum Reply quorum is $F_B+1$, which corresponds to the number of agreeing replies a PBFT client needs to collect from replicas. This quorum is sufficient for safety if nodes sign the block after they have received a PBFT quorum of commit messages, as in our \pbfive protocol. Conversely, in order to ensure safety, simpler consensus protocols such as PBFT and \pbthree must use a Reply quorum that is equal to the Byzantine quorum, since the block signatures are sent together with commit messages.
\subsection{The client}
\label{sec:pbft-client}
In our setting, \new{the PBFT client is just a single virtual entity}: the participant network.
%
This implies a set of changes (simplifications, for the most part) to the duties and operations of our PBFT client.

The original PBFT relies on a client to send request messages to the primary (and, if needed, to other replicas), in order to invoke operations on the replicated state machine.
Once the operation is executed, replicas send back the result directly to the client.
Clients are assumed to be trusted, since the PBFT safety property is insufficient to protect against faulty clients. 
The original PBFT request message is $\langle \textit{REQUEST}, o, t, c \rangle _{\sigma_c}$, where $o$ is the state machine operation, $t$ is a request timestamp, and $c$ is a client identifier.

In our protocol, we instantiate the client and its requests as follows. 
The state machine has a single operation, that appends a new block. The content of the block (e.g., the set of transactions) represents a form of non-determinism and its value will be selected by the primary in the pre-prepare phase.
For these reasons, we omit the operation in the request message, since it is always an implicit ``append'', and the client identifier, since it is unique.

The participants network expects a new valid block to be mined every $\tau$ seconds on average, where $\tau$ is called \textit{target block time}. Being $T_0$ an initial timestamp, called \textit{genesis block timestamp}, that is known in advance by all the replicas, we calculate the request message timestamps by adding $T_0$ to multiples of the desired block time.
In lights of these considerations, all replicas know in advance all valid request messages, that have the form $ \langle \textit{REQUEST}, T_0 + n\tau \rangle$: $n \in \mathbb{N}^+ $, where $n$ is the block height for all blocks (except the genesis one, which is fixed and not mined via the consensus algorithm). Allowing replicas to self-generate the requests allows to not introduce any special role of trusted client.

Non-faulty replicas will generate a request for block $n$ when it is expected to be mined, i.e., when their local clock value exceeds the nominal timestamp of the expected block.

The original PBFT reply message is $\langle \textit{REPLY}, v, t, c, i, r, \rangle _{\sigma_c}$, where $v$ is the current view number, $t$ is the timestamp of the corresponding request, $i$ is the replica identifier, and $r$ is the result of the operation. In our protocol, a block mined at a given height is the result of the append operation of the corresponding request.
Given that valid blocks are broadcast to the participants network once a quorum of signatures by replicas is achieved, we omit the reply messages from our specialized protocol, and replace it with the Bitcoin block propagation mechanism to the participants network.
%
\subsection{Normal operation (no faulty primary)}
\label{sec:pbft-normal-operation}
\begin{figure}
\centering
\includegraphics[width=1\columnwidth]{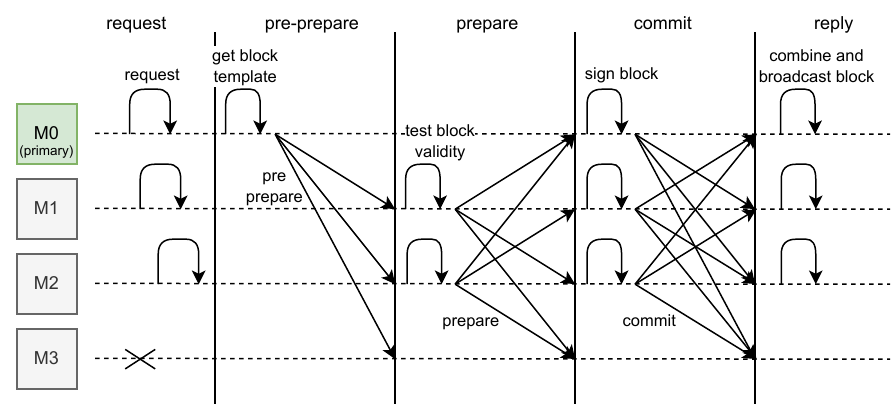}
\caption{Mining network normal operation with $N=4$ nodes, $M_0$ as primary, and $M_3$ as faulty backup.}
\label{fig:pbft-normal-operation}
\end{figure}

We describe the PBFT normal case operations of the mining network, with a special focus on the modifications that allow miners to create, sign and propagate a new valid block. Figure~\ref{fig:pbft-normal-operation} shows the normal case operations with a non-primary faulty replica.

The process starts with a request to append a new block, self-generated by the primary. The operation is non-deterministic, as its result depends on the content of the block to append.
As suggested in~\cite{Castro1999}, we make sure that the primary selects such content independently, and concatenates it with the associated request.

In the pre-prepare phase, the primary $M_0$ gathers a set of transactions from its mempool and forms a template for the next block to be appended. The primary assigns a sequence number $n$ to the block, that corresponds to the height at which the block is expected to be added, then includes the block in the pre-prepare message and broadcasts it to backups for signature\footnote{A faulty primary might send the same, invalid block to all replicas. Therefore, replicas must be able to assess---independently and deterministically---whether the value is correct (and what to do if it is not) based on their current value of the state.}.

A backup (i.e., $M_1$, $M_2$, or $ M_3$ in figure) accepts a pre-prepare message it is valid according to the PBFT rules, its request has been already generated locally by the replica, its timestamp is not in the future according to the local clock of the replica, and the proposed block is also valid. The block template validity is checked by the participant node which is co-located with the replica, and prevents invalid blocks from being signed.
If the request or the block is invalid, the replica ignores the pre-prepare message. This may happen if, e.g., primary clock and the replica's one are not synchronised, and their difference is beyond what is allowed by the request generation rules.
The replica checks the original PBFT conditions, in order to prevent different blocks from being signed at the same height. %
If a backup accepts the pre-prepare message, then it enters the prepare phase and broadcasts the prepare message to all other replicas.

A replica (primary or backup) accepts a prepare message if all the standard PBFT conditions~\cite{Castro1999} are met; no additional checks are present at this stage. A block is said to be \textit{prepared} at replica $i$ in view $v$ and height $n$ iff replica $i$ has received a pre-prepare proposal to append block in view $v$ at height $n$ from the primary, and $Q-1$ backups have acknowledged the proposal, where $Q$ is a Byzantine quorum. The pre-prepare and prepare phases of the algorithm guarantee that non-faulty replicas agree on the block height within a view. When replicas reach an agreement on a block and its height, they proceed to the commit phase, in which they actually sign the prepared block.

In the commit phase, a replica (primary or backup) signs a block, includes the signature in the commit message, and broadcasts the message to other replicas. The addition of the signature to the commit message is a difference with respect to the original PBFT. In \ref{sec:frost-pbft} we will detail the signing phase is articulated in order to generate a confidential quorum certificate using FROST.
A replica accepts a commit message if it contains a valid signature for its corresponding block, and the PBFT conditions are met.

A block is \textit{locally committed} at replica $i$ in view $v$ at height $n$ iff replica $i$ has accepted a Byzantine quorum of commit messages, possibly including its own.
In a naive PBFT implementation, the quorum of commit messages contains a valid Reply quorum of block signatures, and is sufficient to assemble a valid block solution, which can be accepted by the participants network.
In other words, when a quorum of commit messages is collected, a replica can concatenate the signatures from the commit messages, and append the combined signatures to the block solution, in the coinbase transaction;
if the replica is also synchronized with the blockchain, it can subsequently broadcast the new block to the participant network.
It is possible that different replicas broadcast the same block (in terms of transactions), at the same height, but with different sets of signatures, because they received commit messages from different replicas. %
This does not matter for the safety guarantee of the blockchain, and does not cause a blockchain fork or a  reorganization, because block signatures, even if placed in the coinbase transaction, are excluded from the calculation of the root of the transactions Merkle tree, as described in Section~\ref{sec:variations-from-bitcoin-block}.

The PBFT invariants guarantee that if a new block is locally committed at a non-faulty replica, then it is propagated to the network and will eventually be received by all non-faulty participants.
\vspace{-4pt}
\subsection{Checkpoints}
\label{sec:pbft-checkpoints}
The PBFT checkpoint is a mechanism used to discard messages from the log. It is used to guarantee the correctness of the service state that is synchronised among replicas, even when messages have been already discarded from the log.
The original PBFT checkpoint message is $\langle \textit{CHECKPOINT}, n, val_i, lastrep_i,$ $lastrept_i, i \rangle _{\sigma_i}$, where $n$ is the sequence number, $val_i$ is the state machine value, $lastrep_i$ and $lastrept_i$ are the content of the reply message and the timestamp of respective request,  $i$ is a client identifier, and the message is signed by the replica.
A checkpoint is said to be \textit{stable} when its content has been signed by a quorum of different replicas, and the signatures are the proof of correctness of the checkpoint.

In our protocol, we rely on the Bitcoin block propagation mechanism of the underlying participants network for the checkpoint propagation among replicas.
The checkpoint value $n$ is the block height, while the $val_i$ is the whole blockchain content, that we summarize as its tip (or best block hash), and the reply message is the blockchain block at height $n$.
Each block appended to the blockchain is by design agreed upon by a Byzantine quorum of replicas, and the block solution represents the proof of correctness for the checkpoint.
This implies that all checkpoints are stable by design, i.e., they come with a proof that they are the result of the execution of requests by a quorum of replicas.
For this reason, there is no equivalent for an unstable checkpoint in our algorithm: Each replica maintains a single copy of the service state, i.e., the one resulting from the last stable checkpoint, or the last block.
Moreover, the propagation of checkpoints happens via the participants peer-to-peer Bitcoin network.
Each replica, upon receiving a new signed block at height $n$ from the participants, including the ones propagated by the replica itself, does the following: (i) discards all pre-prepare, prepare, and commit messages with sequence number less than or equal to $n$; and (ii) updates the PBFT parameters (namely, low and high watermarks) to signal the replica is ready to append a new block at height $n+1$.
%
\vspace{-5pt}
\subsection{View change}
\label{sec:pbft-view-change}
The view-change protocol provides Liveness (R3) by allowing the mining network to make progress when the primary fails.
In our protocol, as in PBFT, view changes are triggered by timeouts that prevent backups from waiting indefinitely for new blocks.
Our view change is not different from the PBFT one, with the exception that request messages are self-generated by replicas.

Each replica self-generates requests for each block, when these blocks are expected. A backup starts a timer when it self-generates a request and the timer is not already running.
A backup is waiting for a request if it self-generated a request and has not executed it.
It stops the timer when it is no longer waiting to execute the request, but restarts it if at that point it is waiting to execute some other request.
If the timer of backup $i$ expires in view $v$, the backup starts a view change to move the system to view $v+1$.

The same considerations on view-change timeout values that are necessary to achieve Liveness in the original PBFT apply here: the timeout is set to grow faster than network delays across successive views, in particular the timeout grows exponentially with $v$. Moreover, like in PBFT, in order to ensure Safety, if there is a prepared block in view $v$ then the view change protocol propagates the block and its prepared sequence number to $v+1$.

%% file: sec_frost.tex
\section{Certified Byzantine Consensus}
\label{sec:frost}


We hereby address the problem of signing blocks in an aggregated and confidential manner. Moving from a threshold signature scheme known as FROST (Section~\ref{sec:frost-basics}), we design a novel protocol that combines aggregate signatures with a BFT consensus protocol (Section~\ref{sec:frost-pbft}). 
%
%

\subsection{The FROST Signature Scheme}
\label{sec:frost-basics}

A $(k,n)$ \emph{threshold signature} scheme, with $ k \leq n $, requires that at least $ k $ participants over $ n $ cooperate to create a valid signature, i.e., it is not possible to create a valid signature with less than $ k $ participants.
%
%
FROST is a threshold signature scheme that leverages the additive property of Schnorr signatures to quickly combine signatures into an aggregated one~\cite{Komlo2020}. 
The FROST signature scheme defines three main protocols: 
(i) a {\em key generation protocol} that creates secret shares for participants as well as public keys for signature verification;
(ii) a {\em commitment protocol} that creates nonce/commitment share pairs for all participants; these commitments allow to prevent known forgery and replay attacks; 
(iii) a {\em signature protocol} coordinates the generation of the aggregated signature by signers.
We briefly introduce these protocols, whose complete definition can be found in the original work~\cite{Komlo2020}.
%
%

Each participant $M_i$ has a unique identifier $m_i \in \{1, \dots, n\}$. 
Let $\mathbb{G}$ be a group of prime order $ q $ in which the Decisional Diffie-Hellman problem is hard,
$ g $ be a generator of $ \mathbb{G} $, 
and let $ H_1 $ and $ H_2 $ be cryptographic hash functions mapping to $\mathbb{Z}^*_q$. 
We denote by $x \leftarrow A$ that $x $ is selected uniformly randomly from set $ A $. 

\textit{Key Generation.}
Before signing any block, participants 
need to define secret and public keys. 
They share the same cipher suite that specifies the underlying prime-order group details and cryptographic hash function. 
%
The KeyGen protocol consists of two rounds.
%
Afterwards, each participant $M_i$, with $i \in \{1, \dots, n\} $, owns a secret share $s_i$, a public verification share $Y_i = g^{s_i}$, and the group's public key $ Y $.
The public verification share $ Y_i $ allows others to verify the participant signature shares; the group's public key $ Y $ enables the aggregate threshold signature verification, which depends on the set of participants $ n $ and the configured threshold $ k $.

\textit{Commitment.}
In the commitment protocol, participants generate (secret) nonces for signatures and exchange their public commitments, which allow verifying the correct use of nonces.
%
Each participant $M_i$, $i \in \{1, \dots, n\}$, generates a pair of nonces $(d_{i}, e_{i}) \leftarrow \mathbb{Z}^*_q \times \mathbb{Z}^*_q$ and derives the public commitment shares $ (D_{i}, E_{i}) = (g^{d_{i}}, g^{e_{i}})$. 
%
%
%

\textit{Aggregated Signature.}
The aggregate signature protocol works in two rounds. First, each participant generates his signature share. Then, all participants' shares are combined to obtain the final signature. 
Let $ S $ be the set of participants in the signing process; the cardinality of $ S $ is $ \alpha $,  with $ k \leq \alpha \leq n $.
Let $ L = \langle (l, D_l, E_l) \rangle_{l = 1}^\alpha$ be the list of $ \alpha $ participants' commitments.  
When $M_i$ receives the message to sign $ m $, 
%
he can use his secret share $ s_i $ and $ L $ to compute his signature share $z_i$, which can then be sent to all other participants. 
Formally, $M_i$ computes the set of binding values $\rho_l = H_1 (l,m,L)$, $l\in \{1, \dots, \alpha\}$, and derives the group commitment $R = \prod_{l = 1}^{\alpha} D_l \cdot (E_l)^{\rho_l}$ and the challenge $c = H_2(R,Y,m)$. 
Then, $M_i$ computes his signature share on $ m $ as $z_i = d_i + (e_i \cdot \rho_i) + \lambda_i \cdot s_i \cdot c$, using $(d_i,e_i)$ corresponding to $(i,D_i,E_i) \in L$, and $ S $ to determine the $i$-th Lagrange coefficient $\lambda_i$.
%
Since nonces cannot be used multiple times, $M_i$ deletes the $\left( (d_i,D_i), (e_i,E_i) \right)$ pair from his local storage.
Then, $M_i$ sends $z_i$ to every other participant in $S$.

The second round starts when $M_i$ receives all other signature shares $z_l$. 
%
For verification, $M_i$ checks if the equality $g^{z_l} = R_l \cdot Y_l ^{c\cdot \lambda_l}$ holds for each received $z_l$. 
If the verification is successful, $M_i$ aggregates the signature shares locally by computing $z = \sum_{i \in S} z_i $. 
The resulting aggregated signature of $ m $ is 
$\sigma = (R,z)$, that can be verified as single-party signature.

%
\subsection{FROSTing PBFT}
\label{sec:frost-pbft}
The challenge of combining FROST with PBFT arises from the presence of Byzantine nodes that may refuse to sign blocks.
A na\"ive integration of FROST would not work in practice due to the difficulty of defining upfront the quorum of participants $ S $ that will collaborate to compute the aggregated signature $ z $.
Integrating FROST with BFT requires to review the commitment and aggregate signature protocols (Section~\ref{sec:frost-basics}). We need rules for exchanging $(D_i,E_i)$ pairs and the set of signers $ S $, two critical information for determining the binding values $ \rho_i $, the challenge $ c $, and Lagrange coefficients $ \lambda_i $ that reconstruct the secret used to sign messages. 


Two protocols are presented, called \pbthree{} and \pbfive{}.
The first represents the trivial way to use FROST in an asynchronous setting, and optimizes the communication by shipping potential signature shares with the PBFT commit messages. Unfortunately, this solution requires an exponential number of potential commitments and signature shares and can be used only when the set of signers is small.
The second introduces new rounds at the end of PBFT consensus, which allow to minimize the information exchanged to produce a valid block signature. To guarantee the protocol liveness in presence of Byzantine nodes that may refuse to sign, this protocol uses the idea proposed by ROAST~\cite{ruffing2022roast}.
Both protocols enhance nodes with the cryptographic primitives presented in Section~\ref{sec:frost-basics}.
Either way, at the end of the protocol, each participant has a block $ m $ with the related signature $\sigma = (R,z)$. The integrity of $ m $ can be validated using the traditional Schnorr verification algorithm~\cite{Schnorr1989}.


\subsubsection{{\em\pbthree{}} (3-Phase Frosted-BFT)}
\begin{figure}
    \centering
    \includegraphics[width=\columnwidth]{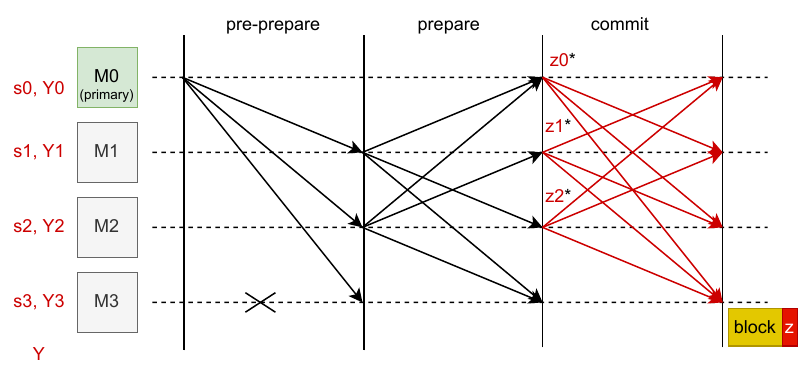}
    \caption{Normal operation (no faulty primary) of \pbthree{}. Replicas exchange a set of potential signature shares in the commit phase. When a replica receives these signature shares combines them to reconstruct the aggregate signature for the block in a decentralized manner.}
    \label{fig:pbft-frost-3rounds}
\end{figure}
This protocol aims to optimize the number of rounds by allowing replicas to directly exchange multiple, potential signature shares. The replicas themselves will figure out autonomously how to combine the signature shares to compute the correct aggregated signature for the block. 
Unlike \pbfive{}, this protocol decouples the commitment protocol from the PBFT consensus.
%
\new{
Each replica generates nonces and exchanges information with other replicas to let them derive the public commitments needed to sign blocks in \pbthree{}. This commitment protocol uses a hierarchical deterministic key derivation~\cite{fornaro2018elliptic, Pieter2019}, which allows determining $(D_i,E_i) $ for each participant $ i $ starting from an extended public commitment of $ i $ and locally available information.
From here on, we assume each participant knows its own nonces and the public commitments of all replicas.
}
The number of commitments to generate considers that, for each block request, a participant will exchange $ \gamma $ signature shares with all other participants.
%
The $\gamma $ parameter is defined observing that determining the set of block signers $ S $ in advance is not possible, due to the presence of Byzantine replicas. However, computing an aggregated signature requires only $ k $ signature shares, with the threshold equal to the Signet quorum size. 
In this case, the Signet quorum size equals the PBFT quorum size $ Q $, because signature aggregation takes place in the commit phase that, in turn, waits for $ Q $ messages as indicated in Section~\ref{sec:pbft-quorum-size}.
Each participant $M_i$ determines all $k$-combinations of $ n $ known participants $ \mathcal{S} = \langle S^{(j)} \rangle_{j = 1}^\gamma $, where $ \gamma = \binom{n}{k} $ is the number of combinations. 
%
Since the number of signature shares $\gamma$ grows almost as $O(2^n)$, this protocol is feasible only when $ n $ is small, e.g., in a small mining network.
The nonces and public commitments are indexed by the participant identifier, the PBFT sequence number (i.e., block height), and 
the identifier of a specific combination of signers (i.e., $ j \in \{1,\dots,\gamma\}$).
In this way, a participant will use a different nonce for each generated signature share.
Moreover, each participant can readily retrieve the correct nonce/commitment pairs for each combination of signers in a non-interactive manner.

In \pbthree{}, the signature aggregation protocol is executed at once with the PBFT consensus protocol.
The \pbthree{} protocol has indeed the same number of phases as the traditional PBFT. Commit messages are extended to transfer also signature shares. 
As usual, we assume that each replica of the consensus protocol participates in the signing process. 
%
Figure~\ref{fig:pbft-frost-3rounds} shows \pbthree{}, highlighting the modified messages in red.
The pre-prepare and prepare phases of \pbthree{} exactly match the phases from Section~\ref{sec:pbft}.
%
%
When the commit phase starts, $ M_i $ determines the list of public commitments for every combination of $ k $ participants in $ \mathcal{S} $ that include $M_i$.
%
$M_i$ computes a signature share $z_i^{(j)}$ for each of these combinations, obtaining $ \mathcal{Z}_i = \langle z_i^{(j)}\rangle_{j = 1}^\gamma$, which is then sent to all other replicas via the commit message.  
When the quorum in the commit phase is reached, each replica has all the information to aggregate signature shares of others and create a valid block certificate.
Each participant uses the received commit messages to identify a set of $ k $ participants, whose index is $j^*$, and accordingly extract the signature share $ z_i^{(j^*)} $ from $ \mathcal{Z}_i $ for each $ i \in S^{(j^*)} $. 
So, he retrieves all $ k $ participants' public commitments, indexed by $ j^* $, 
and reconstructs $ R $ (as per Section~\ref{sec:frost-basics}).
To complete \pbthree{}, participants aggregate the signature shares, $ z = \sum_i z_i^{(j^*)} $, and save $ \sigma = (R, z) $ as the certificate of block $ m $. 
%

%
%
%
%

\subsubsection{{\em\pbfive{}} (5-Phase Frosted-BFT)}
This variant introduces two main changes to the BFT protocol (Section~\ref{sec:pbft}). First, it blends the commitment protocol and the aggregate signature protocol into the (normal case) PBFT. Second, it extends PBFT with additional rounds to guarantee liveness in case Byzantine nodes play the role of signers.
We assume that each replica of the consensus protocol is a participant in the signing process. 
\begin{figure}
    \centering
    \includegraphics[width=\columnwidth]{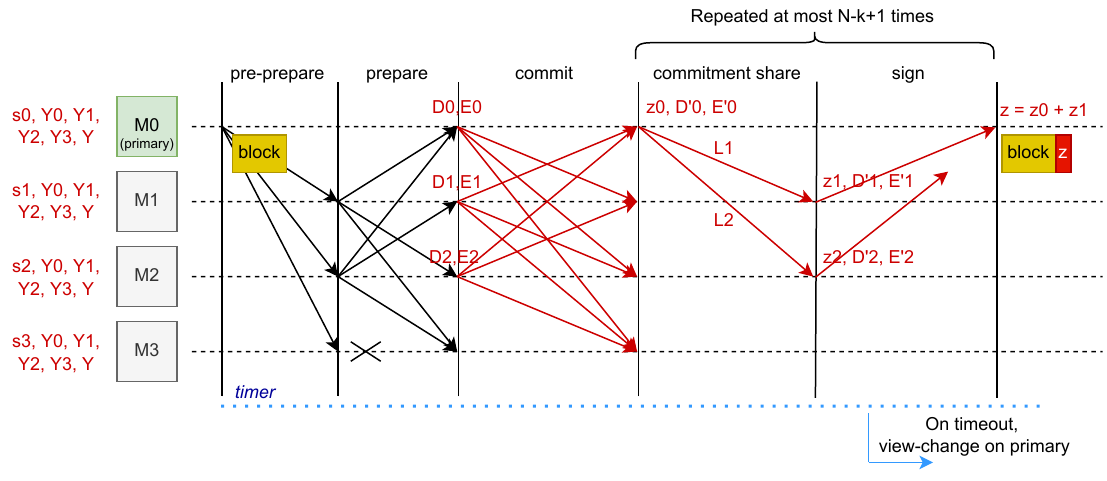}
    \caption{Normal operation (no faulty primary) of \pbfive{}. Replicas exchange their public nonce commitments in the commit phase. In the commitment share phase, the primary defines the $ L $ parameter enabling the creation of partial signatures. The latter are exchanged in the sign phase, enabling replicas to aggregate the signature in a decentralized manner.}
    \label{fig:pbft-frost-5rounds}
\end{figure}
Replicas collaborate to apply the threshold signature on the block agreed upon consensus. 
As per Figure~\ref{fig:pbft-frost-5rounds}, \pbfive{} introduces new rounds in PBFT, namely commitment-share and sign. The commitment-share and sign rounds are executed according to ROAST (RObust ASynchronous Threshold signatures), which wraps the FROST protocol as described in \cite{ruffing2022roast}.
When a message is prepared by a replica $ i $, it runs the {\em commitment protocol} to randomly determine the nonce/commitment share pairs $ \left((d_i,e_i), (D_{i}, E_{i})\right) $. Public commitments $ (D_{i}, E_{i}) $ are piggybacked to the commit message and exchanged with other replicas leveraging the PBFT protocol. 
The primary holds a list of responsive signers, among which the set of candidate signers will be defined.
Replicas that send the public commitments in their commit message are considered as part of the initial set of responsive signers, and will be considered as candidate for the a signing session in the commitment-share phase.

After replicas exchange the commit messages, a set of {\em commitment-share} phases of \pbfive{} takes place. 
The primary defines a set of signers $ S $ among active replicas; the selection policies for defining the set $ S $ follow the rules of ROAST. $ S $ has cardinality $\alpha$ with $\alpha = k$, including the primary itself, and we require $ k = F_B + 1$ (see Section~\ref{sec:pbft-quorum-size}); in such a configuration, at least a honest signer is included, thus preventing forgery of aggregate signatures over invalid blocks. 
%
Even though the primary might exclude nodes suspected to be potentially unresponsive, malicious nodes are unknown a priori, so they could still be included in $ S $. For this reason the primary can initiate multiple and concurrent commitment-share sessions, maintaining a set of responsive signers, i.e., signers that have responded to all previous signing requests. As soon as there are at least $k$ responsive signers in the set, the primary will initiate a new commitment-share session.
When the primary determines $ S $, he creates and sends the list of signers' public commitment $ L = \langle (l, D_l, E_l) \rangle_{l \in S}$ to other replicas. 
Knowing $ L $ (and, consequently, $ S $), other replicas can determine all the information to compute the signature share on the block, namely $ \rho_l $, $ \lambda_l $, and $ c $, for $ l \in S$.

The {\em sign} phase of \pbfive{} allows replicas to run the {aggregate signature protocol} presented in Section~\ref{sec:frost-basics}. 
They create and exchange with the primary the signature shares $z_i$, with $ i \in \{1, \dots, \alpha\}$, together with a new public commitment to be possibly used in another commitment-share session. If any signature share $z_i $ is not valid, the primary marks the replica as malicious, so that it will not be included in subsequent commitment-share phases.
When the primary receives all other signature shares $ z_i $, with $ i \in S $, it can combine them to derive the aggregate signature $ z = \sum_i z_i $. The aggregate signature $ \sigma = (R, z) $ certifies the block $ m $. When the aggregate signature is correctly defined, the primary broadcasts the block and complete \pbfive{}.
As demonstrated in ROAST, the commitment-share sessions will eventually finish, and a non-faulty primary will receive all the signatures, in at most $N-k+1$ commitment-share sessions, under the hypothesis that the number of possible backup replica failures $F_B + F_C$ is at most $N-k$.
The view-change protocol described in Section~\ref{sec:pbft-view-change} allows to provide liveness also in presence of a faulty primary, which delays (but does not compromise) the ROAST protocol.
When a view change is triggered by the block timeout, the (possibly new) primary replica will act as a new semi-trusted coordinator, that will run again the aggregate signature protocol.
Note that the view-change cannot change values the quorum has agreed upon, so the block content cannot be updated. 
It is also worth pointing out that, thanks to the properties of threshold signatures, even if the set of signers $ S $ and related parameters change, a valid aggregate signature will be produced. This property follows from Lagrange interpolation: A different set of participants basically provides a different set of points over the same $t-1$ degree polynomial, where the signature secret lies.

\subsection{Proof sketches for \pbfive{}}
We now informally argue that the \pbfive{} algorithm satisfies the safety and liveness properties.
The safety property relies on the usual cryptographic assumptions and a threshold adversary model with threshold $N>3f$, while the liveness additionally relies on the partial synchrony of the network (as in PBFT).

\begin{theorem}[Safety]
If a set of replicas produced a valid signature for block $b_1$ with sequence number $n$ in view $v$, then no valid signature will be produced for block $b_2$ with $n$ and $v$ by another set of replicas.
\end{theorem}
\begin{proofsk}
By contradiction, assume that for view $v$ and sequence number $n$,
there are two blocks $b_1$ and $b_2$, with $b_1\neq b_2$, for which $f+1$ signature shares have been collected. Consider non-faulty replicas $r_1$ and $r_2$ that signed for $b_1$ and $b_2$, respectively. If $r_1=r_2$, we already get a contradiction: a correct replica signed two different blocks for $ n $ and $ v $. If $r_1\neq r_2$, by the safety property of PBFT, there cannot be two correct replicas that commit to two different blocks for the same sequence number. Similarly, the claim also holds for $v'>v$, since PBFT guarantees that
if at least a correct replica locally committed the block $b$ in $v$, which is the precondition to sign $b$, then no other request will be considered for the same sequence number $n$ in later views $v'>v$.
\end{proofsk}

\begin{theorem}[Liveness]
All valid block proposal are eventually committed and signed by all correct replicas.
\end{theorem}
\begin{proofsk}
The claim follows from the proof of termination of ROAST (Theorem 4.3 of~\cite{ruffing2022roast}), and by the liveness property of PBFT, by taking care of triggering view-changes if a replica detects a Byzantine aggregator of signature shares.
\end{proofsk}
The requirements of Correctness and Calmness are satisfied by construction, as a pre-prepare message is accepted only if its block is valid according to the protocol rules (Correctness), its request has been generated locally by the replica, and its timestamp is not in the future (Calmness). Finally, the Confidentiality requirement is guaranteed by the confidentiality of the signature created using FROST.

%% file: sec_evaluation.tex
\section{Evaluation}
\label{sec:evaluation}

\subsection{Geographically distributed environment}
We evaluate the performance of \pbfive{} in a geographically distributed environment, involving up to $22$ mining nodes, placed in $8$ different European regions of Amazon Web Services (AWS).\footnote{Namely: eu-west-1 (Ireland), eu-central-1 (Frankfurt), eu-south-1 (Milan), eu-north-1 (Stockholm), eu-west-2 (London), eu-central-2 (Zurich), eu-south-2 (Spain), eu-west-3 (Paris). We assign a sequential identifier to each node and determine the AWS region where to deploy it using the modulo function.}
According to data collected by cloudping in 2022\footnote{\url{https://www.cloudping.co/grid/p_50/timeframe/1Y}}, the median latency between these regions ranges between $20$~ms and $50$~ms, while the intra-region median latency stays below $4$~ms. 

The main measure of performance we consider is \textit{consensus latency} (or just latency, for short), which represents the time needed by the mining network to reach an agreement and sign a new block. Measured at the primary node, it is the difference between the time at which a new signed block is submitted to the participant network (end of the consensus) and the time at which a new block is proposed to the mining network with a \textit{pre-prepare} message (beginning of the consensus algorithm).

\begin{figure}
    \centering
    \includegraphics[width=1\columnwidth]{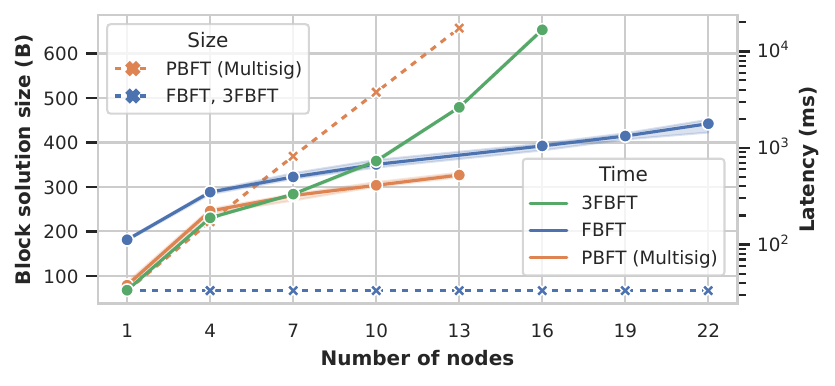}
    \caption{Solution size and average consensus latency achievable with na\"ive PBFT, \pbfive{}, and \pbthree{}.}
    \label{fig:exp:latency-signatureoverhead-geographically}
\end{figure}

In Fig~\ref{fig:exp:latency-signatureoverhead-geographically}, we compare \pbfive{} with the two baseline variants, the na\"ive PBFT which produces a block solution as the concatenation of signatures by a threshold of the mining nodes, and \pbthree{}.
For the three variants, Fig~\ref{fig:exp:latency-signatureoverhead-geographically} shows the size of the block solution (i.e., the block signature), which represents a witness of the mining network agreement and is broadcast to the participant network together with the block itself.
PBFT produces a block solution whose size increases with the mining network size (from $222$~bytes with $4$ replicas to $657$~bytes with $13$ replicas). 
Notably, the \texttt{OP\_CHECKMULTISIG} opcode, used by participants to verify the block solution, allows checking at most $15$ public keys.
Both \pbthree{} and \pbfive{} represent an improvement over the PBFT baseline because they use FROST for creating a quorum certificate and produce a single Schnorr signature, which can be verified with an ad-hoc Taproot output. Therefore, the block solution size is $ 67$~bytes, no matter the number of miners.
Fig~\ref{fig:exp:latency-signatureoverhead-geographically} also compares the three algorithms in terms of latency for different sizes of the mining network, here in absence of load. The experiments confirm that \pbfive{} shows higher latency than PBFT, which is motivated by the presence of additional rounds needed for the FROST signature aggregation. Moreover, \pbthree{} shows lower latency than \pbfive{} in small mining networks, but its performances degrade very quickly when the number of nodes is greater than $ 10 $. This is motivated by the calculation of all possible combinations of a \textit{Byzantine quorum} of signatures out of all the possible signers, leading to a prohibitively high latency of $19.2$~s with $ 16 $ nodes.
With $ 22 $ mining nodes, \pbfive{} registers an average latency of 
$1.7$~s
in a setting spread across $ 8 $ AWS regions. This value of consensus latency is largely below our requirement of a new block every $ 60 $~s.
\begin{figure}
    \centering
    \includegraphics[width=1\columnwidth]{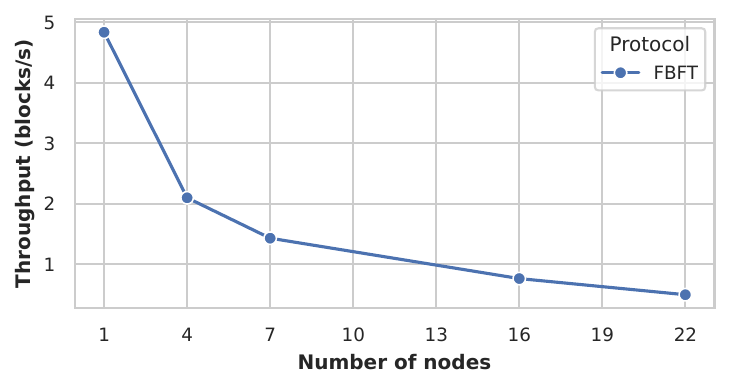} 
    \caption{Average throughput without load}
    \label{fig:exp:fbft:thr}
\end{figure}
\begin{figure}
    \centering
    \includegraphics[width=1\columnwidth]{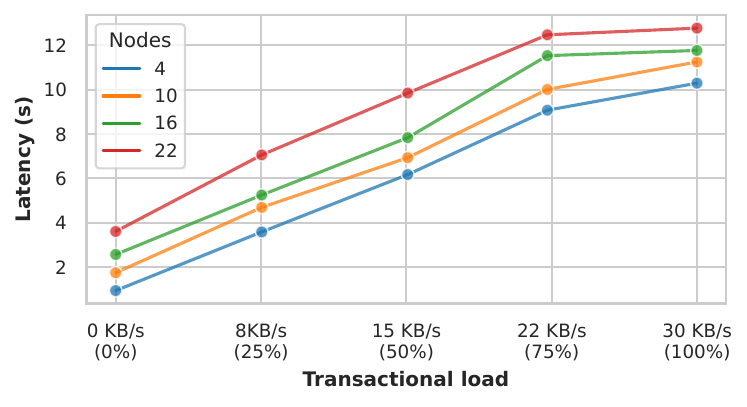} 
    \caption{Average \pbfive{} latency under different load conditions}
    \label{fig:exp:fbft:lat}
\end{figure}
\begin{figure}
    \centering
    \includegraphics[width=1\columnwidth]{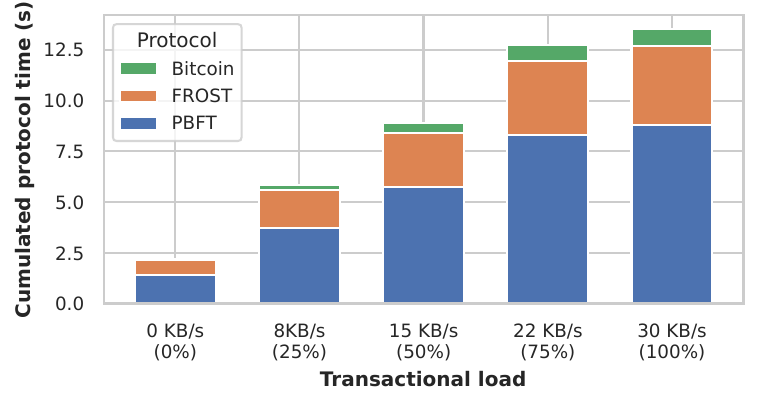} 
    \caption{Avg time spent by \pbfive{} in its protocols, 16 nodes}
    \label{fig:exp:fbft:time}
\end{figure}
\begin{figure}
    \centering
    \includegraphics[width=1\columnwidth]{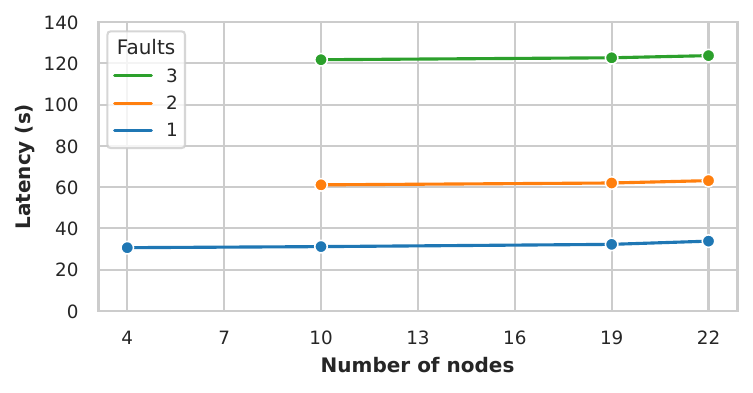}
    \caption{Evaluation of \pbfive{}: throughput, time spent in its protocols, and consensus latency.}
    \label{fig:exp:fbft:fail}
\end{figure}

Fig.~\ref{fig:exp:fbft:thr} reports the maximum throughput achievable with \pbfive{} across the AWS European regions. The throughput is the number of blocks that could be produced by the mining network per unit of time. It is slightly lower than the inverse of the consensus latency, because it also takes into account the time for block propagation in the mining network. We configure the experiments so that the network produces blocks as fast as possible (i.e., we ``disable'' the calmness, forcing miners to recover a blockchain with a genesis block time in the past).
As expected, the throughput decreases as the mining network size increases. 
This is mainly due to the quadratic communication complexity of \pbfive{}, which builds on PBFT.

In Fig~\ref{fig:exp:fbft:lat} we investigate the impact of incoming transactions on consensus latency. 
We configure the mining network to generate a new block every minute in the steady state, during which we evaluate the performances. However, since we set the genesis block timestamp 30 minutes in the past, there is a warm-up period in which the miners will try to mine the first 30 blocks at the highest rate achievable with the given network conditions.
In order to generate load for the mining network, we set up additional $8$ participant nodes that submit transactions to the mining network. The warm-up period described above allows the clients to fill their wallets with a number of coins sufficient to generate a transaction load at the desired rate. The client rate is set to target a given block size that we describe as $0\%, 25\%, 50\%, 75\%, 100\%$ of maximum block size ($1.8$~MB).
As Fig~\ref{fig:exp:fbft:lat} shows, when the load increases, the consensus latency increases as well. Indeed, the increase of the block size slows down the exchange of the \textit{pre-prepare} message by the primary, and the block validity check by backups. We experience that the impact of transaction load is linear for all the mining network configurations. Nevertheless, even with blocks at full load and $22$ mining nodes, the maximum latency we experienced is around $13$~s, below our requirement to mine a block every $60$~s.

To agree and sign the next block, \pbfive{} uses different protocols: PBFT for consensus, FROST for signature aggregation, and Bitcoin for block validation. Fig.~\ref{fig:exp:fbft:time} details the time spent in these different stages, under different load conditions, with $16$ mining nodes.
When the load is $ 0 $~KB/s, \pbfive{} closes empty blocks in $2170$~ms: it spends $64\%$ of time to complete the PBFT phases; $35.1\%$ of time to exchange and aggregate the signature shares, whereas the Bitcoin block validity and submission checks take less than $1\%$ of the time. 
If the load is at its maximum, then PBFT takes $65\%$ of time, the signature aggregation takes $29\%$, and the block validity check takes $6\%$ of time.

Finally, we evaluate the block latency in presence of failures. When the primary appears as faulty, \pbfive{} uses the view change protocol to elect a new primary and recovers from failure.
Fig.~\ref{fig:exp:fbft:fail} shows the impact of failures in the worst-case scenario, where the primaries of subsequent views fail consecutively, and multiple view changes are triggered before finding the agreement on the next block.
After $60$~s, we forcefully terminate the primary of the mining network in the initial view, and possibly up to two other primaries expected for the next views. 
We set the initial view change timeout to $30$~s for an expected block time of $60$~s.
Almost for every mining network size, the consensus latency is strongly delayed by the view-change protocol, which doubles subsequent timeouts with the number of subsequent views. It increases from less than $ 2$~s 
to $\approx 30$~s, with $ 1 $ failure, 
to $\approx 60$~s, with $ 2 $ concurrent failures, 
to $\approx 120$~s, with $ 3 $ concurrent failures ($123.7$~s with 22 nodes). 
When the view change is completed, the consensus protocol recovers the delayed blocks at the maximum throughput, and continues to mine with calmness at a consensus latency that is less than $ 2$~s.

Overall, it appears that \pbfive{} can provide Byzantine fault tolerance, network confidentiality, and efficient usage of block solution space, for just a reasonable increment in consensus latency.

%% file: sec_related.tex
\section{Related work}
\label{sec:related}
This work draws on ideas from three research areas: permissioned DLTs, fault tolerant consensus algorithms, and signature schemes.
\vspace{-2pt}
\subsection{Permissioned DLTs}

Even if Bitcoin is primarily designed for the public network, there exist previous examples of Bitcoin-like ledger meant for private networks. In
particular, Elements\footnote{\url{https://blockstream.com/elements}}, whose production deployment---the ``Liquid'' sidechain~\cite{nick2020liquid}---uses a BFT consensus algorithm within a permissioned mining network consisting of cryptocurrency businesses\footnote{\url{https://help.blockstream.com/hc/en-us/articles/900003013143-What-is-the-Liquid-Federation-}}.
Elements provides additional features with respect to  Bitcoin, including confidential assets~\cite{poelstra2018confidential} and more advanced programming capabilities, which may be further expanded~\cite{o2017simplicity}.

Elements is the closest piece of previous work in terms of technologies and \new{Bitcoin reuse goals}. However, to the best of our knowledge, no public specification for its proprietary BFT approach exists, and the open sourced components\footnote{\url{https://github.com/ElementsProject}} do not include its implementation.


The second largest DLT born public and then adapted to permissioned settings is Ethereum~\cite{wood2014ethereum}. For example, an Ethereum-like ledger designed for private networks is Hyperledger Besu\footnote{\url{https://www.hyperledger.org/use/besu}}, which supports a PoA consensus based on Istambul BFT~\cite{Moniz2020}. An implementation is available in open source\footnote{\url{https://github.com/hyperledger/besu/tree/main/consensus/ibft}}.
Another Ethereum-like ledger designed for private networks is Concord\footnote{\url{https://blogs.vmware.com/opensource/2018/08/28/meet-project-concord}}.
This is possibly the closest work to ours, in spirit, but (i) it implements SBFT~\cite{GolanGueta2019} instead of PBFT as a consensus algorithm (a quality open source implementation exists\footnote{\url{https://github.com/vmware/concord-bft}}); (ii) it works with BLS signatures instead of Schnorr to generate a quorum certificate; and (iii) it has Ethereum instead of Bitcoin as a foundation.

Point (iii) is a profound differentiator, for the reasons we discuss in Section~\ref{sec:bitcoin-strenghts}. In particular, with respect to the ``focus'', ``reliability'', and ``extensibility'' dimensions, it is worth noting that Ethereum exhibits a Turing-complete language as a key feature, and focuses on the development of complex decentralized applications via smart contracts, more than on digital payments.
Among the relevant applications of its smart contracts, we find the issuance of cryptocurrency tokens~\cite{vogelsteller2015eip}, NFTs \cite{entriken2018eip}, the creation of financial businesses that do not rely on intermediaries (e.g., decentralized exchanges~\cite{angeris2019analysis}, DeFi applications). \new{Several of these decentralized applications have experienced attacks by cybercriminals~\cite{su2021evil}, not rarely executed by leveraging the subtleties of the scripting language}.



There is a host of other relevant permissioned DLTs, whose main difference with respect to our approach is the absence---by design---of any attempt to profit from existing code bases and software from major public blockchains: They are custom DLTs, that often redesign/reimplement much from scratch.
Notable examples are:

\begin{itemize}[leftmargin=*]
    \item \emph{Hyperledger Fabric}\footnote{\url{https://www.hyperledger.org/use/fabric}}, a general purpose DLT that enables the development of enterprise applications, not necessarily financial; it is designed particularly for private networks. Among its components, there is a BFT consensus module based on BFT-SMaRT~\cite{bessani2014state}, but its development appears to have been stopped\footnote{\url{https://github.com/bft-smart/fabric-orderingservice}}.
    \item \emph{Corda}\footnote{\url{https://www.corda.net}}, a DLT designed for the financial industry. It has a token-based data model (UTXO) like Bitcoin, and a Turing-complete programming language like Ethereum. Its main applications are the issuing of digital assets, or currencies, payments, and global trade. It has a pluggable consensus mechanism: Its ``\emph{notaries}'' can run either a crash fault tolerant (CFT) consensus, such as RAFT, or a BFT consensus, like PBFT. Neither the BFT specification nor its implementation are available in the open source repository\footnote{\url{https://github.com/corda}}, and apparently no aggregated signature scheme is included.
    \item \emph{Hyperledger Sawtooth}\footnote{\url{https://www.hyperledger.org/use/sawtooth}} allows to deploy private DLT networks with a variety of consensus algorithms, including PBFT and Proof of Elapsed Time~\cite{chen2017security}, a recent proposal which uses a trusted execution engine to securely generate a random waiting time and then choose a node with the smallest waiting. Sawtooth has an open source implementation\footnote{\url{https://github.com/hyperledger/sawtooth-pbft}}, with an incomplete PBFT implementation and no FROST-like signature aggregation.
    \item \emph{Diem}\footnote{\url{https://developers.diem.com/docs/welcome-to-diem}} (formerly Libra) is a novel DLT platform which implements a Turing-complete programming language designed for safe and verifiable transaction-oriented computation. It employs a custom BFT algorithm called DiemBFT~\cite{baudet2019state} based on Hotstuff~\cite{Yin2019}, which has an open source implementation\footnote{\url{https://github.com/diem/diem/tree/main/consensus}}.
\end{itemize}

Finally, there is Hamilton~\cite{lovejoy2022high}, which is close to our approach in terms of use cases: It is a DLT designed to support payments in a private network. It exhibits a token-based data model, and it has an open source implementation\footnote{\url{https://github.com/mit-dci/opencbdc-tx}}. It inherits certain elements from Bitcoin (such as the UTXO model and even the elliptic curve used in all cryptographic primitives). Relevant differences: its ledger is not a blockchain; its consensus protocol is not Byzantine but CFT; its ledger is meant to stay private; its main focus is on obtaining transactional scalability on-ledger.
\input{tab_permissioned}

\vspace{-2pt}
\subsection{Byzantine Fault Tolerance}
\label{sec:related-pbft}
Lamport et al.~\cite{Lamport1982} firstly introduced the problem of a distributed system reaching agreement in the presence of Byzantine failures.
Different surveys review the most relevant BFT consensus protocols (e.g.,~\cite{Garay2020,Sankar2017,Xiao2020,Alsunaidi2019,Pahlajani2019}).
PBFT by Castro and Liskov~\cite{Castro1999,Castro2002} is considered the reference solution for practical implementations. 
%

In PBFT, replicas exchange messages using an all-to-all communication pattern, hence PBFT does not scale well; attempts to optimize it exist, along different directions, such as communication pattern (e.g.,~\cite{GolanGueta2019,Yin2019}), leader rotation (e.g.,~\cite{Abraham2019,Buchman2016,DiemTeam2021,Veronese2009}), view-change optimization (e.g.,~\cite{Buchman2016,Buterin2019}),  pipelining (e.g.,~\cite{Buterin2019, Yin2019}), and speculative/optimistic execution (e.g.,~\cite{Abraham2017,Aublin2015,GolanGueta2019,Kotla:2010,Pass2018,DiemTeam2021}). 


The adoption of ``collectors'' reduces the number of exchanged messages, obtaining a linear communication pattern. A collector is a designated replica (usually the leader) that receives and broadcasts messages to all the other replicas.
Based on this idea, SBFT~\cite{GolanGueta2019} uses a dual-mode protocol with an optimistic fast path (when replicas are mostly in sync) and a fallback slow path (more PBFT-like).
%
%
%
This dual-mode protocol increases complexity in favor of performance.

To avoid dual-mode, HotStuff~\cite{Yin2019} uses a collector in combination with threshold signatures to generate a quorum certificate for each protocol phase.
This is the most closely related work to ours. 
By using a single collector, HotStuff increases the protocol complexity to avoid weakening its robustness, to face cases when, e.g., the collector himself is a Byzantine node.
 
DiemBFT~\cite{DiemTeam2021} enhances HotStuff to improve throughput. Using a Pacemaker mechanism, whose design is however not fully specified, DiemBFT synchronizes the consensus phases
to simplify the consensus protocol. However, DiemBFT, just like HotStuff, does not aggregate quorum signatures in a fully decentralized manner.

In PBFT,
the leader changes
only when a problem is detected. 
As an alternative, the rotating leader strategy~\cite{Veronese2009} proposes to change the leader after every attempt to commit (e.g.,~\cite{Buchman2016,Yin2019}).
To efficiently solve the leader selection problem, deterministic as well as more sophisticated policies have been proposed (e.g.,~\cite{Abraham2019,DiemTeam2021}).
%
We postpone as future work the design of a more sophisticated consensus policy with better fairness properties (cfr. Section~\ref{sec:future}).


As we have discussed, it was the advent of blockchains to renew the interested in consensus algorithms (e.g.,~\cite{Buchman2016,Buterin2019,DiemTeam2021,Gilad2017,Moniz2020}).
%
%
Notably, PoS~\cite{King2012} was quickly proposed as an alternative to PoW, known to have energy consumption issues, especially for permissionless blockchains (e.g.,~\cite{Bashir2020,Bentov2016,Buterin2019}).
Research on PoS led to two main approaches: 
One proposes a chain-based PoS that mimics PoW (e.g.,~\cite{Bentov2016,Kiayias2017}); 
the other proposes a BFT-based PoS, where randomly selected nodes participate in a multi-round voting protocol to determine the next blocks to append (e.g.,~\cite{Buterin2019,Gilad2017,Song2019}).
BFT-inspired protocols are preferred due to their deterministic block finality.

Other notable work in this area include: Istanbul BFT~\cite{Moniz2020}, a variant of PBFT tailored for blockchains;
Pass and Shi~\cite{Pass2018} work on a consensus protocol for permissioned and permissionless blockchains that combines a fast path and a slow one (as SBFT does);
Algorand~\cite{Gilad2017} and its proposal to scale PBFT by seeking consensus only on a subset of two-thirds of nodes, selected using PoS. 
%
These works do not consider quorum certificates or aggregated block signatures.
Byzcoin \cite{byzcoin} combines the use of PoW with BFT protocols to realize highly-performant open consensus protocols.

In contrast to all these scalability-oriented works, our setting considers a limited number of miners and expresses the calmness requirement (P4). Therefore, we seek protocols that favor simplicity and robustness, and focus on their tight integration with Bitcoin.
\vspace{-2pt}
\subsection{Threshold signatures}
%

In threshold signature schemes,
at least $ k $ participants over $ n $, with $ k \leq n $, collaborate for generating a valid signature {\em on-behalf} of the group.
%
%
%
%
%
Shoup~\cite{Shoup2000} defined one of the most used threshold signature schemes, based on RSA (e.g.,~\cite{Cachin2005,GolanGueta2019,Yin2019,Thai2019}). 
It requires a trusted, centralized dealer for key generation, and then uses non-interactive signature share generation and signature verification protocols.

Unfortunately, both the RSA signature scheme and the trusted dealer make this solution unsuited to our Bitcoin-derived setting.

%
%
%
%
%
%
%
Gennaro et al.~\cite{Gennaro2001} propose a threshold DSA signature scheme, with $ k < n / 2$,  where a trusted centralized dealer is adopted. 
%
In~\cite{Gennaro2016}, Gennaro et al. propose a dealer-less approach supporting the case $ k < n $. 
However, DKG is costly and impractical. 
Then, Gennaro and Goldfeder~\cite{Gennaro2018,Gennaro2020} presented 
an ECDSA-based protocol 
supporting efficient DKG, 
obtaining faster signing than~\cite{Gennaro2016} 
and requiring less data to be transmitted.
%
%
%
In a closely related work, Lindell et al.~\cite{Lindell2018} propose an efficient threshold ECDSA scheme, which employs different methods to neutralize any adversarial behavior.
%

A detailed (and more extensive) review of threshold ECDSA schemes can be found in~\cite{Aumasson2020}.
Although ECDSA is fast and secure, aggregated signatures
cannot be easily obtained with it, so we avoided this route.
%
%

Conversely, BLS~\cite{Boneh2004} and Schnorr~\cite{Schnorr1989} schemes can be easily transformed into threshold versions by supporting the sum of partial signatures with no overhead~\cite{Ergezer2020}.\footnote{Schnorr and BLS signature schemes natively work with elliptic curves. While signature verification in BLS is more computationally demanding than ECDSA and Schnorr, signature generation in BLS is completely deterministic (thus preventing tampering).}
%
%
%
%
%
In particular, Boldyreva~\cite{Boldyreva2002} proposed the most widely adopted approach for threshold BLS signatures.
DKG does not require a trusted dealer, and the signature generation does not require participant interaction (or any zero-knowledge proof). It can only tolerate up to $ k < n/2 $ malicious parties,
but it allows to periodically renew the secret shares. 

Most BFT protocols with a collector use threshold BLS signatures (e.g.,~\cite{GolanGueta2019,DiemTeam2021,Yin2019}).
%
%
Recently, Tomescu et al.~\cite{Tomescu2020} 
proposed a more efficient BLS signature scheme, that improves signing and verification time. 
Threshold BLS signature schemes rely on pairing-based cryptography~\cite{Boneh2004}; 
%
However, this may be challenging to implement in practice in our platform, since BLS signatures are not supported in Bitcoin. 
%
%

Schnorr signatures received increased interest recently, and they have been included in the Bitcoin protocol.
%
%
%
Komlo and Goldberg~\cite{Komlo2020} propose FROST,
an efficient Schnorr-based threshold scheme, whereby signing can be performed in two rounds, or optimized to a single round with preprocessing.
%
%
FROST is currently considered the most efficient scheme
for generating threshold Schnorr signatures~\cite{Crites2021}, and is the one we have adopted in the present work. Ruffing et al.~\cite{ruffing2022roast} propose ROAST (RObust ASynchronous Schnorr Threshold Signatures), a wrapper protocol around FROST that provides liveness guarantees in presence of malicious nodes and asynchronous networks. The \pbfive{} protocol we propose considers the idea introduced by ROAST to guarantee liveness while aggregating signature shares in the case of Byzantine signers. 


%% file: tab_permissioned.tex
\begin{table}
\caption{Comparison of major permissioned distributed ledgers}\label{tab:related-permissioned}
\adjustbox{max width=\columnwidth}{%
\begin{tabular}{l|ccccc}
& Byzantine 
& Confidential 
& Distributed 
& Schnorr  
& Open  
\\ 
& Fault Tolerance 
& Quorum Certificate 
& Ledger Technology 
& Quorum Certificate 
& Source 

\\ \hline \hline
Elements
& Strong Federations~\cite{Dilley16strongfederation}
& Unclear                 
& Bitcoin 
& Unclear                      
& Partially

\\ \hline
Hyperledger Besu   
& IstanbulBFT\cite{Moniz2020}   
& No            
& Ethereum
& No  
& Yes       

\\ \hline
Concord 
& SBFT~\cite{GolanGueta2019}  
& Yes, BLS\cite{Boneh2004}  
& Ethereum 
& No                      
& Yes       

\\ \hline  
Hyperledger Fabric
& No, Raft~\cite{Ongaro14raft}
& No
& Fabric
& No   
& Yes

\\ \hline  
Corda  
& Unclear
& Unclear  
& Corda
& Unclear  
& Partially

\\ \hline  
Hyperledger Sawtooth  
& PBFT~\cite{Castro1999}          
& No 
& Sawtooth 
& No           
& Yes

\\ \hline  
Diem      
& DiemBFT\cite{baudet2019state} 
& No     
& Diem
& No        
& Yes

\\ \hline  
Project Hamilton
& No, Raft~\cite{Ongaro14raft}
& No                   
& Hamilton       
& No        
& Yes      

\\ \hline  
Our solution  
& \pbfive{} 
& Yes                   
& Bitcoin       
& Yes, FROST~\cite{Komlo2020}        
& Yes

\\ \hline  
\end{tabular}
}
\end{table}

%% file: sec_conclusions.tex
\section{Future work}
\label{sec:future}
We are working on two primary research topics, both of which aim at making our prototype closer to a deployable, real-world solution: One concerns the improvement of the core PoA consensus algorithm itself (``\emph{Dynamic federation}'' and ``\emph{Improved fairness}'' sections); the second one involves programming the blockchain to generate layer-2 constructions that contribute the missing features we expect of actual retail payment systems (``\emph{Privacy and Scalability}'').
\paragraph{Dynamic federation}
Our mining network configuration is currently static: Miners are assigned their role at the beginning and cannot be changed without recreating the whole blockchain. This is unacceptable for real-world use, since dynamic reconfigurations are bound to happen for a variety of reasons: e.g., a member may need to change its public key periodically; a new member may need to join the federation; members may be removed.
One hypothetical solution is to represent voting rights as non-fungible tokens (NFT) in our very chain: Block validity conditions would be the same as the spending condition of an NFT on the chain, and a transfer of the NFT would represent a change in the mining configuration.



\paragraph{Improved fairness}
When BFT algorithms are used in blockchains, full fairness\footnote{We use the Chain Quality property definition from \cite{GKL15,GKL20}, with parameters $\mu \in \mathbb{R}$ and $l \in \mathbb{N}$, which states that, for any honest party $P$ with chain $C$, it holds that, for any consecutive $l$ blocks of $C$, the ratio of honest blocks is at least $ \mu $. In our model, ideal fairness requires the Chain Quality property to hold for $\mu =1-F_B/N$ (ideal chain quality). Note that we do not need to include crash failures since they cannot contribute negatively to the chain quality.} is not achieved unless the primary can generate at most a single block before a view change. This would prevent a Byzantine primary to censor transactions. A fair mining network needs to rotate the leader at each block, e.g., using an election technique based on cryptographic sortition via Verifiable Random Functions from Algorand~\cite{gilad2017algorand}.
%
%


\paragraph{Privacy and Scalability}
Our prospective model assumes that most of the privacy and scalability issues of actual retail payments are solved at the $2^{nd}$ layer. In particular we are experimenting with a dedicated Payment Channel Network
similar to the Bitcoin Lightning~\cite{poon2016bitcoin}, but with a less spontaneous topology, one that better fits our permissioned scenario and that is programmed over the distributed engine presented in this paper.
Some assessments of, e.g., the scalability of Lightning Network already exist~\cite{tikhomirov2020quantitative}; however, to get the whole picture we need to consider how different factors interplay: the raw throughput of the network, its business-driven topology, the level of privacy achieved by participants, and the costs of locking liquidity into channels by the routing intermediaries.

%


\section{Conclusions}
\label{sec:conclusions}

We presented a Bitcoin-like, permissioned, distributed ledger in which valid blocks are signed by a federation of trusted actors and transactions enjoy deterministic finality.
Block signatures are aggregated via a threshold scheme based on FROST, that preserves the confidentiality of the \new{mining} configuration and quorum.
We showed how such a federation, via PBFT, could operate correctly also under Byzantine failures of a subset of the nodes.

Our design embodies one way of inheriting all the algorithms, data structures, and software of Bitcoin---but its PoW-based consensus protocol---in order to make its full technological stack \new{openly} available to permissioned settings managed by trusted actors.

What for? 
Bitcoin has inspired innumerable other blockchains and is perhaps the closest thing we have to an open standard for payments in the ``crypto domain''. It is possible that its technological stack---constantly scrutinized, improved, evolved\footnote{As an example of the forward-thinking surrounding the eldest blockchain, consider that work exists suggesting how to transform its cryptographic apparatus into a quantum-resistant one, even on-the-fly during a quantum attack, via a soft fork~\cite{stewart2018committing}.
}---will one day percolate\footnote{\new{This perspective has famed historical precedents. 
It is not unlike reusing the exact same technological stack from a decentralized, public network (the Internet) into private, ``permissioned'' networks (intranets): TCP/IP. At the dawn of the networking era, the idea of a convergent public/private stack was unheard of, and a host of custom, proprietary networking suites were deployed for ``permissioned'' use cases. But eventually, the good-enough and widely adopted TCP/IP won over most specialized (and mutually incompatible) protocols. It became a \emph{de facto} standard.
We are a very long way from a similar turn of events in the realm of digital payment systems. And, it may well be argued that a shared technological ground is unlikely to ever materialize. Still, we proved the idea makes technical sense, and we would better be ready.}} into the blockchain-friendly portion of the banking and financial ecosystem for old and new use cases. Solutions such as the one we present here pave the way for such a possibility.